\documentclass[preprint,12pt]{revtex4-1}
\usepackage{slashed,enumerate}

\begin{document}

\title{Some chirality-related properties of the 4-D massive Dirac propagator and determinant in an arbitrary gauge field}
\author{Jin Hur} \email{hurjin@kias.re.kr}
\affiliation{School of Computational Sciences, Korea Institute for Advanced Study, Seoul 130-012, Korea}
\author{Choonkyu Lee} \email{cklee@phya.snu.ac.kr}
\affiliation{Department of Physics and Astronomy and Center for Theoretical Physics\\ Seoul National University, Seoul 151-742, Korea}
\author{Hyunsoo Min} \email{hsmin@dirac.uos.ac.kr}
\affiliation{Department of Physics, University of Seoul, Seoul 130-743, Korea}

\begin{abstract}
For a 4-D massive Dirac field in the background of arbitrary gauge fields, we show that the Dirac propagator and functional determinant are completely determined by knowledge of the corresponding
quantities for just one of the chirality sectors of the second-order Dirac
operator.
This generalizes the related, previously known, statements in (anti-)self-dual background gauge fields. The logarithms of the (renormalized) functional determinants from the two chirality sectors are shown to be different only by a term reflecting the integrated chiral anomaly.
\end{abstract}
\maketitle

\section{Introduction}
In 4-D Euclidean spacetime it has been known for some time that, for a spin-$\frac{1}{2}$ Dirac field defined in the background of an (anti-)self-dual gauge field, there exist simple formulas connecting the associated particle propagator and functional determinant to those of a scalar (i.e., spin-0) field \cite{bccl,brownlee,thooft,browncreamer,pac}. Explicitly, in the background of (Abelian or non-Abelian) gauge fields $A_\mu$ satisfying the self-duality condition $F_{\mu\nu} = \,^*F_{\mu\nu}$ (here, $F_{\mu\nu} = \partial_\mu A_\nu - \partial_\nu A_\mu - i [A_\mu,A_\nu]$ and $\,^*F_{\mu\nu} \equiv \frac{1}{2} \epsilon_{\mu\nu\lambda\delta} F_{\lambda\delta}$), let $S$ denote the massive spin-$\frac{1}{2}$ propagator defined by
\begin{eqnarray}
\left( \slashed{D}+im \right) S = I, \label{diracpropagatordef}
\end{eqnarray}
where $\slashed{D} \equiv \gamma_\mu D_\mu$ and $D_\mu \equiv \partial_\mu - iA_\mu$ (with our $\gamma$-matrices satisfying $\{\gamma_\mu,\gamma_\nu\} = -2\delta_{\mu\nu}$), and $\triangle$ denotes the corresponding spin-0 propagator defined by
\begin{eqnarray}
\left( -D^2+m^2 \right) \triangle = I, \qquad \left( D^2 \equiv D_\mu D_\mu \right). \label{triangledef}
\end{eqnarray}
Then, $S$ is expressed in terms of $\triangle$ by the formula \cite{brownlee}
\begin{eqnarray}
S = \left( \slashed{D}-im \right) \triangle \left( \frac{1+\gamma_5}{2} \right) + \triangle \slashed{D} \left( \frac{1-\gamma_5}{2} \right) + \frac{1}{im} \left( 1-\slashed{D}\triangle\slashed{D} \right) \left( \frac{1-\gamma_5}{2} \right). \label{Sform}
\end{eqnarray}
This connection on propagators can be utilized to derive the relations between the corresponding functional determinants or one-loop effective actions. That is, if $\Gamma_{\text{ren}}^F (A;m)$ denotes the renormalized fermion one-loop effective action (using the `minimal' subtraction) derived from the bare effective action
\begin{eqnarray}
\Gamma^F(A;m) &\sim& -\ln \det \left[ -i\slashed{D}+m \right] + \text{const.} \nonumber\\
&\sim& -\frac{1}{2} \ln \det \left[ \slashed{D}\slashed{D}+m^2 \right] + \text{const.}, \label{bareeffectiveaction}
\end{eqnarray}
it can be related to the corresponding scalar effective action $\Gamma_{\text{ren}}^S(A;m)$ (again using the minimal subtraction) by \cite{pac}
\begin{eqnarray}
\Gamma_{\text{ren}}^F(A;m) = -2\Gamma_{\text{ren}}^S(A;m) - \frac{1}{2} n_F \ln \left(\frac{m^2}{\mu^2}\right), \label{fermionscalarrelation}
\end{eqnarray}
where $\mu$ is the renormalization scale and $n_F$ denotes the number of zero eigenmodes for the related massless Dirac operator. The formulas (\ref{Sform}) and (\ref{fermionscalarrelation}) generalize the corresponding findings for the massless case, first stated in Refs. \cite{bccl,thooft}.

Recently, making use of the connection (\ref{fermionscalarrelation}) in an important way, we (with G. Dunne) were able to determine the fermion or quark contribution to the one-loop effective action in a single instanton background, for any value of the quark mass \cite{insdet}. As a rationale for this simple relationship existing between the amplitudes involving different spin fields (actually including even vector or spin-1 fluctuation fields), one refers to the supersymmetry of the system \cite{thooft,zumino} for which the self-dual nature of background gauge fields is supposed to be important. Then, with a general non-self-dual background gauge fields, one would not expect that there exist appropriate generalizations of the above formulas, which reduce to (\ref{Sform}) and (\ref{fermionscalarrelation}) if gauge fields are taken to be self-dual. But, as for the Dirac field (with a non-chiral gauge coupling), we show in this paper that such generalizations do exist, except that the scalar-field-related amplitudes need to be replaced, in the general case, by the amplitudes defined in any particular chirality sector of the second-order Dirac operator, $\slashed{D}\slashed{D}+m^2$. These are true for both the propagator and the one-loop effective action. To state any relation between the effective actions, one must be precise about the renormalization procedure for them --- we assume minimal, mass-independent, subtraction terms throughout (so that the mass dependence of the effective action does not get mingled with renormalization effects). The new relationships, being not specific to self-dual backgrounds, have also obvious parallels in the Minkowski space. Our observation should be useful in future studies of fermion effective actions in nontrivial backgrounds.

This paper is organized as follows. In Sec \ref{sec2} we derive our new formulas generalizing the above relations (\ref{Sform}) and (\ref{fermionscalarrelation}) to the case of a massive Dirac field in an arbitrary background gauge field. In deriving the formula for the one-loop effective action, the well-known chiral anomaly \cite{adler} in an integrated form enters our discussion. Then, in Sec. \ref{sec3}, we illustrate these results explicitly for the case of gauge fields corresponding to constant field strengths. Sec. \ref{sec4} contains some concluding comments. There are also two appendices containing some supplementary materials.

\section{Dirac propagators and determinants in arbitrary gauge fields} \label{sec2}
We will first present our formula for the Dirac propagator $S$, defined by (\ref{diracpropagatordef}). [We use the operator notations of Refs. \cite{bccl,brownlee}, and $A_\mu(x) \equiv A_\mu^a(x) T^a$ when $T^a$ denote appropriate group generator matrices]. In our discussion the spacetime signature does not matter much, but for definiteness we will continue to assume the Euclidean signature. If we here introduce the second-order Dirac propagator $\mathcal{G}$ by
\begin{eqnarray}
\left[ \slashed{D}\slashed{D}+m^2 \right] \mathcal{G} = 1, \label{Geq}
\end{eqnarray}
$S$ can be identified with
\begin{eqnarray}
S = \left( \slashed{D}-im \right) \mathcal{G} = \mathcal{G} \left( \slashed{D}-im \right). \label{SG}
\end{eqnarray}
Then, on resolving the identity by positive and negative chiral projections
\begin{eqnarray}
1 = \left(\frac{1+\gamma_5}{2}\right) + \left(\frac{1-\gamma_5}{2}\right),
\end{eqnarray}
we can  write
\begin{eqnarray}
\mathcal{G} = \mathcal{G}^{(+)} \left(\frac{1+\gamma_5}{2}\right) + \mathcal{G}^{(-)} \left(\frac{1-\gamma_5}{2}\right). \label{Gsplit}
\end{eqnarray}
On the other hand, we obtain from (\ref{SG}) the relation
\begin{eqnarray}
\slashed{D}\slashed{D}\mathcal{G} = \slashed{D}\mathcal{G}\slashed{D}
\end{eqnarray}
and hence, combining this with (\ref{Geq}), the following identity should be true \cite{brownlee}:
\begin{eqnarray}
\mathcal{G} = \frac{1}{m^2} \left[ 1-\slashed{D}\mathcal{G}\slashed{D} \right].
\end{eqnarray}
This then leads to
\begin{eqnarray}
\mathcal{G} \left(\frac{1\pm\gamma_5}{2}\right) = \mathcal{G}^{(\pm)} \left(\frac{1\pm\gamma_5}{2}\right) &=& \frac{1}{m^2} \left[ 1-\slashed{D}\mathcal{G}\slashed{D} \right] \left(\frac{1\pm\gamma_5}{2}\right) \nonumber\\
&=& \frac{1}{m^2} \left[ 1-\slashed{D}\mathcal{G}^{(\mp)}\slashed{D} \right] \left(\frac{1\pm\gamma_5}{2}\right), \label{Ggamma}
\end{eqnarray}
since $\slashed{D} (\frac{1\pm\gamma_5}{2}) = (\frac{1\mp\gamma_5}{2}) \slashed{D}$. Based on (\ref{Gsplit}) and (\ref{Ggamma}), it is thus possible to express $\mathcal{G}$ by either
\begin{eqnarray}
\mathcal{G} = \mathcal{G}^{(+)} \left(\frac{1+\gamma_5}{2}\right) + \frac{1}{m^2} \left[ 1-\slashed{D}\mathcal{G}^{(+)}\slashed{D} \right] \left(\frac{1-\gamma_5}{2}\right) \label{G1}
\end{eqnarray}
or
\begin{eqnarray}
\mathcal{G} = \frac{1}{m^2} \left[ 1-\slashed{D}\mathcal{G}^{(-)}\slashed{D} \right] \left(\frac{1-\gamma_5}{2}\right) + \mathcal{G}^{(-)} \left(\frac{1-\gamma_5}{2}\right). \label{G2}
\end{eqnarray}
[In the $m \to 0$ limit this expression will become singular whenever there exist normalizable zero eigenmodes for $\slashed{D}$ \cite{brownlee}; in this work, $m$ is kept at nonzero value to avoid any subtlety concerning the strictly massless case].

Thanks to (\ref{G1}) or (\ref{G2}), the full second-order Dirac propagator $\mathcal{G}$ can be constructed from the knowledge on either one of $\mathcal{G}^{(\pm)}$. For the latter object, it will be convenient to adopt the chiral representation for $\gamma$-matrices, i.e.,
\begin{eqnarray}
\gamma _{\mu }=\left(
\begin{array}{cc}
 0 & \sigma _{\mu } \\
 -\bar{\sigma }_{\mu } & 0
\end{array}
\right), \qquad \text{(with $\sigma _{\mu }=(\vec{\sigma },i)$ and $\bar{\sigma }_{\mu }=(\vec{\sigma },-i)=(\sigma_\mu^\dagger)$)}
\end{eqnarray}
which leads to the diagonal matrix for $\gamma_5$
\begin{eqnarray}
\gamma_5 \equiv \gamma_1\gamma_2\gamma_3\gamma_4 = \left(
\begin{array}{cc}
  1 & 0 \\
  0 & -1
\end{array}
\right)
\end{eqnarray}
For these $\gamma$-matrices the operator $\slashed{D}\slashed{D}+m^2$ takes the form
\begin{eqnarray}
\slashed{D}\slashed{D}+m^2 &=& \left(
\begin{array}{cc}
 -\sigma \cdot D \; \bar{\sigma } \cdot D + m^2 & 0 \\
 0 & -\bar{\sigma } \cdot D \; \sigma \cdot D + m^2
\end{array}
\right) \nonumber\\
&=& \left(
\begin{array}{cc}
 -D^2-\frac{1}{2} \eta _{\mu \nu a}^{(-)} \sigma _a F_{\mu \nu } + m^2 & 0 \\
 0 & -D^2-\frac{1}{2} \eta _{\mu \nu a}^{(+)} \sigma _a F_{\mu \nu } + m^2
\end{array}
\right) \label{chiraldecomposition}
\end{eqnarray}
where $\eta _{\mu \nu a}^{(\pm)}$ are 't Hooft symbols \cite{thooft}, introduced through the relations like
\begin{eqnarray}
\bar{\sigma }_{\mu } \sigma _{\nu }=\delta _{\mu \nu }+i \eta _{\mu \nu a}^{(+)} \sigma _a, \quad \sigma _{\mu } \bar{\sigma }_{\nu }=\delta _{\mu \nu }+i \eta _{\mu \nu a}^{(-)} \sigma _a.
\end{eqnarray}
In this chiral representation we can write
\begin{eqnarray}
\mathcal{G}^{(+)} = \left(
\begin{array}{cc}
  \bar{\mathcal{G}}^{(+)} & 0 \\
  0 & \bar{\mathcal{G}}^{(+)}
\end{array}
\right), \quad \mathcal{G}^{(-)} = \left(
\begin{array}{cc}
  \bar{\mathcal{G}}^{(-)} & 0 \\
  0 & \bar{\mathcal{G}}^{(-)}
\end{array}
\right)
\end{eqnarray}
with the reduced propagators $\bar{\mathcal{G}}^{(\pm)}$ ( $2\times2$ matrices in the spin space) satisfying the relation
\begin{eqnarray}
\left( -D^2-\frac{1}{2} \eta _{\mu \nu a}^{(\mp)} \sigma _a F_{\mu \nu } + m^2 \right) \bar{\mathcal{G}}^{(\pm)} = 0. \label{Gpmeq}
\end{eqnarray}
Hence, if the gauge field satisfies the self-duality condition $F_{\mu\nu} = \,^*F_{\mu\nu}$ (which is possible in the Euclidean spacetime), we have $\eta_{\mu\nu a}^{(-)} F_{\mu\nu} = 0$ and so $\bar{\mathcal{G}}^{(+)} = \frac{1}{-D^2+m^2} \equiv \triangle$ (see (\ref{triangledef})); i.e., $\bar{\mathcal{G}}^{(+)}$ coincides with the spin-0 propagator and (\ref{G2}) yields the form previously given in \cite{brownlee}
\begin{eqnarray}
\mathcal{G} = \triangle \left(\frac{1+\gamma_5}{2}\right) + \frac{1}{m^2} \left[ 1-\slashed{D}\triangle\slashed{D} \right] \left(\frac{1-\gamma_5}{2}\right). \label{Gbrownlee}
\end{eqnarray}
Also, in an arbitrary background gauge field, (\ref{Ggamma}) implies that following relationships hold between $\bar{\mathcal{G}}^{(+)}$ and $\bar{\mathcal{G}}^{(-)}$:
\begin{eqnarray}
\bar{\mathcal{G}}^{(+)} = \frac{1}{m^2} \left[ 1 + \sigma \cdot D \bar{\mathcal{G}}^{(-)} \bar{\sigma} \cdot D \right], \qquad \bar{\mathcal{G}}^{(-)} = \frac{1}{m^2} \left[ 1 + \bar{\sigma} \cdot D \bar{\mathcal{G}}^{(+)} \sigma \cdot D \right]. \label{Gpmrelation}
\end{eqnarray}

From (\ref{SG}) and (\ref{G2}) the first-order Dirac propagator $S$ can also be expressed in terms of either one of $\bar{\mathcal{G}}^{(\pm)}$. By making use of the relation
\begin{eqnarray}
\slashed{D} \left[ 1-\slashed{D}\mathcal{G}^{(\pm)}\slashed{D} \right] \left(\frac{1\mp\gamma_5}{2}\right) = m^2 \mathcal{G}^{(\pm)} \slashed{D} \left(\frac{1\mp\gamma_5}{2}\right),
\end{eqnarray}
it is not difficult to show that $S$ is expressed by either
\begin{eqnarray}
S = \left( \slashed{D}-im \right) \mathcal{G}^{(+)} \left(\frac{1+\gamma_5}{2}\right) + \mathcal{G}^{(+)} \slashed{D} \left(\frac{1-\gamma_5}{2}\right) + \frac{1}{im} \left[ 1-\slashed{D}\mathcal{G}^{(+)}\slashed{D} \right] \left(\frac{1-\gamma_5}{2}\right) \label{S1}
\end{eqnarray}
or
\begin{eqnarray}
S = \mathcal{G}^{(-)} \slashed{D} \left(\frac{1+\gamma_5}{2}\right) + \frac{1}{im} \left[ 1-\slashed{D}\mathcal{G}^{(-)}\slashed{D} \right] \left(\frac{1+\gamma_5}{2}\right) + \left( \slashed{D}-im \right) \mathcal{G}^{(-)} \left(\frac{1-\gamma_5}{2}\right). \label{S2}
\end{eqnarray}
Notice that the form (\ref{Sform}) is recovered as a special case of (\ref{S1}), i.e., when the gauge field strength is restricted to be self-dual.

We now tern to the discussion of the one-loop effective action which plays a central role in quantum field theory. The Pauli-Villars regularized form of the bare spinor effective action (\ref{bareeffectiveaction}) can be presented, using the Schwinger proper-time representation \cite{schwinger}, in the form
\begin{eqnarray}
\Gamma_\Lambda^F(A;m) = \frac{1}{2} \int_0^\infty \frac{\mathrm{d}s}{s} \left( e^{-m^2s}-e^{-\Lambda^2s} \right) F(s). \label{properGamma}
\end{eqnarray}
Here,
\begin{eqnarray}
F(s) = \mathrm{Tr} \left[ e^{-s\slashed{D}\slashed{D}}-e^{-s(-\partial^2)} \right] \equiv \int \mathrm{d}^4x\; \mathcal{T}\!\mathit{r} \left\langle x \left| \left[ e^{-s\slashed{D}\slashed{D}}-e^{-s(-\partial^2)} \right] \right| x \right\rangle
\end{eqnarray}
with `$\mathcal{T}\!\mathit{r}$' denoting the trace over both Dirac spinor and internal `isospin' indices. Then the renormalized effective action, in the `minimal' subtraction scheme, can be identified with the expression
\begin{eqnarray}
\Gamma_{\text{ren}}^F(A;m) = \lim_{\Lambda\to\infty} \left\{ \Gamma_\Lambda^F(A;m) - \frac{1}{3} \frac{1}{(4\pi)^2} \ln \frac{\Lambda^2}{\mu^2}  \int \mathrm{d}^4x \;\mathrm{tr}\left( F_{\mu\nu}F_{\mu\nu} \right) \right\}, \label{Gammaren}
\end{eqnarray}
where $\mu$ is the renormalization scale and `$\mathrm{tr}$' represents the trace over isospin indices only. The effective action based on other renormalization prescriptions can be related quite simply to our expression (\ref{Gammaren}), as discussed for instance in Ref. \cite{kwon}.

Now, in view of the chiral decomposition (\ref{chiraldecomposition}) given for the operator $\slashed{D}\slashed{D}+m^2$, one would expect that $\Gamma_{\text{ren}}^F(A;m)$ can be expressed by the sum of the two contributions
\begin{eqnarray}
\Gamma_{\text{ren}}^F(A;m) = \Gamma_{\text{ren}}^{(+)}(A;m) + \Gamma_{\text{ren}}^{(-)}(A;m),
\end{eqnarray}
where $\Gamma_{\text{ren}}^{(\pm)}(A;m)$ denote the (suitably renormalized) effective actions associated with 2-component spinor fields having $\bar{\mathcal{G}}^{(\pm)}$ (see (\ref{Gpmeq})) as their propagators. Including appropriate renormalization counterterms (see Appendix \ref{appendixA}), $\Gamma_{\text{ren}}^{(\pm)}(A;m)$ can be given by the proper-time representations
\begin{eqnarray}
 \Gamma_{\text{ren}}^{(\pm)}(A;m) =&& \lim_{\Lambda\to\infty}\frac{1}{2}
 \left\{ \int_0^\infty \frac{\mathrm{d}s}{s} \left( e^{-m^2s}-e^{-\Lambda^2s} \right) F^{(\pm)}(s) \right. \nonumber\\
&& \left. + \frac{1}{(4\pi)^2}  \ln \frac{\Lambda^2}{\mu^2} \int \mathrm{d}^4x
\;\left(- \frac{1}{3}\mathrm{tr}\left( F_{\mu\nu}F_{\mu\nu} \right) \pm \frac{1}{2}\mathrm{tr}(F_{\mu\nu}\,^*F_{\mu\nu})\right) \right\}, \label{Gammarenpm}
\end{eqnarray}
where we defined
\begin{eqnarray}
F^{(\pm)}(s) = \int \mathrm{d}^4x\; \overline{\mathcal{T}\!\mathit{r}} \left\langle x \left| \left[ e^{-s(-D^2-\frac{1}{2} \eta _{\mu \nu a}^{(\mp)} \sigma _a F_{\mu \nu })} - e^{-s(-\partial^2)} \right] \right| x \right\rangle. \label{Fdef}
\end{eqnarray}
(We denoted the trace over \emph{2-component} spinor and isospin indices by $\overline{\mathcal{T}\!\mathit{r}}$). Given the connections (\ref{Gpmrelation}) for the two propagators $\bar{\mathcal{G}}^{(\pm)}$, it would be natural to suspect the existence of certain connections for the two quantities $\Gamma_{\text{ren}}^{(\pm)}(A;m)$ also. Having that in mind, let us define the quantity
\begin{eqnarray}
\Delta\Gamma(A;m) &=& \Gamma_{\text{ren}}^{(+)}(A;m) - \Gamma_{\text{ren}}^{(-)}(A;m) \nonumber\\
&=& \lim_{\Lambda\to\infty} \frac{1}{2}\left\{\int_0^\infty \frac{\mathrm{d}s}{s} \left( e^{-m^2s}-e^{-\Lambda^2s} \right) \overline{\mathrm{Tr}} \left[ e^{-s(-\sigma \cdot D \; \bar{\sigma } \cdot D)} - e^{-s(-\bar{\sigma} \cdot D \; \sigma \cdot D)} \right] \right. \nonumber\\
&& \qquad \left. + \frac{1}{(4\pi)^2}  \ln \frac{\Lambda^2}{\mu^2} \int \mathrm{d}^4x \;\mathrm{tr}(F_{\mu\nu}\,^*F_{\mu\nu}) \right\}, \label{DeltaGammadef}
\end{eqnarray}
($\overline{\mathrm{Tr}}$ denotes the extension of $\overline{\mathcal{T}\!\mathit{r}}$ to include also the trace over spacetime coordinates), and then it should be possible to write
\begin{eqnarray}
\Gamma_{\text{ren}}^F(A;m) = 2 \Gamma_{\text{ren}}^{(+)}(A;m) - \Delta\Gamma(A;m) \label{GammaDeltaGamma1}
\end{eqnarray}
and also
\begin{eqnarray}
\Gamma_{\text{ren}}^F(A;m) = 2 \Gamma_{\text{ren}}^{(-)}(A;m) + \Delta\Gamma(A;m). \label{GammaDeltaGamma2}
\end{eqnarray}

We can determine the quantity $\Delta\Gamma(A;m)$ generally in an explicit form. [Fry \cite{Fry} studied the related quantity assuming a specific form for the background fields, but did not give an explicit formula (given in (\ref{DeltaGammaresult}) below)]. To that end, write
\begin{eqnarray}
\Delta\Gamma(A;m) = \Delta\Gamma(A;M) - \int_{m^2}^{M^2} \mathrm{d}\bar{m}^2 \frac{\mathrm{d}(\Delta\Gamma(A;\bar{m}))}{\mathrm{d}\bar{m}^2}, \label{DeltaGammamMi}
\end{eqnarray}
taking $M^2$ at some large value. We then note that, on the basis of (\ref{DeltaGammadef}),
\begin{eqnarray}
\frac{\mathrm{d}(\Delta\Gamma(A;\bar{m}))}{\mathrm{d}\bar{m}^2} &=& -\frac{1}{2} \int_0^\infty \mathrm{d}s\; e^{-\bar{m}^2s}\; \overline{\mathrm{Tr}} \left[ e^{-s(-\sigma\cdot D \; \bar{\sigma} \cdot D)} - e^{-s(-\bar{\sigma}\cdot D \; \sigma \cdot D)} \right] \nonumber\\
&=& -\frac{1}{2} \overline{\mathrm{Tr}} \left[ \frac{1}{-\sigma \cdot D \; \bar{\sigma} \cdot D + \bar{m}^2} - \frac{1}{-\bar{\sigma} \cdot D \; \sigma \cdot D + \bar{m}^2} \right]. \label{DeltaGammap}
\end{eqnarray}
On the other hand, from (\ref{chiraldecomposition}),
\begin{eqnarray}
&& \bar{m}^2\; \overline{\mathrm{Tr}} \left[ \frac{1}{-\sigma \cdot D \; \bar{\sigma} \cdot D + \bar{m}^2} - \frac{1}{-\bar{\sigma} \cdot D \; \sigma \cdot D + \bar{m}^2} \right] \nonumber\\
&&\qquad\qquad = \mathrm{Tr} \left[ \frac{\bar{m}^2}{\slashed{D}\slashed{D}+\bar{m}^2} \left( \frac{1+\gamma_5}{2} \right) - \frac{\bar{m}^2}{\slashed{D}\slashed{D}+\bar{m}^2} \left( \frac{1-\gamma_5}{2} \right) \right] \nonumber\\
&&\qquad\qquad = \mathrm{Tr} \left[ \frac{\bar{m}^2}{\slashed{D}\slashed{D}+\bar{m}^2} \gamma_5 \right],
\end{eqnarray}
and therefore (\ref{DeltaGammap}) can be rewritten as
\begin{eqnarray}
\bar{m}^2 \frac{\mathrm{d}(\Delta\Gamma(A;\bar{m}))}{\mathrm{d}\bar{m}^2} = -\frac{1}{2}\; \mathrm{Tr} \left[ \frac{\bar{m}^2}{\slashed{D}\slashed{D}+\bar{m}^2} \gamma_5 \right]. \label{mDeltaGammap}
\end{eqnarray}
What we have in the right hand side of (\ref{mDeltaGammap}) is the familiar quantity in the so-called index calculation \cite{bcl,weinberg}. As was asserted in \cite{bcl}, this quantity -- essentially the spacetime integral of the vacuum expectation value $\bar{m} \langle \psi^\dagger \gamma_5 \psi \rangle$ of a Dirac field of mass $\bar{m}$ -- should be independent of $\bar{m}$ and has in fact the value proportional to the integrated form of the famous chiral anomaly \cite{adler} for the divergence of the axial vector current $\psi^\dagger \gamma_5 \gamma_\mu \psi$:
\begin{eqnarray}
\mathrm{Tr} \left[ \frac{\bar{m}^2}{\slashed{D}\slashed{D}+\bar{m}^2} \gamma_5 \right] = -\frac{1}{(4\pi)^2} \int \mathrm{d}^4x\; \mathrm{tr}(F_{\mu\nu}\,^*F_{\mu\nu}). \label{chiralanomaly}
\end{eqnarray}
[The derivation of this result based on the Schwinger proper-time method is also given in Appendix \ref{appendixB}]. Using the result (\ref{chiralanomaly}) in (\ref{DeltaGammamMi}), we obtain
\begin{eqnarray}
\Delta\Gamma(A;m) = \Delta\Gamma(A;M) - \frac{1}{2} \frac{1}{(4\pi)^2}\ln \frac{M^2}{m^2}
\int \mathrm{d}^4x\; \mathrm{tr}(F_{\mu\nu}\,^*F_{\mu\nu}). \label{DeltaGammamM}
\end{eqnarray}
Moreover, the expression for $\Delta\Gamma(A;M)$, when $M$ is sufficiently large, is readily found using our representation (\ref{DeltaGammadef}): using the heat kernel expansion given in Appendix \ref{appendixA}, its value for large $M$ is
\begin{eqnarray}
\Delta\Gamma(A;M) &=& \lim_{\Lambda\to\infty} \frac{1}{2}\frac{1}{(4\pi)^2} \left\{  \int_0^\infty \frac{\mathrm{d}s}{s} \left( e^{-M^2s}-e^{-\Lambda^2s} \right)  \left[ - \int \mathrm{d}^4x\; \mathrm{tr} (F_{\mu\nu}\,^*F_{\mu\nu}) + O(s) \right] \right. \nonumber\\
&&\qquad \qquad\left. +  \ln \frac{\Lambda^2}{\mu^2} \int \mathrm{d}^4x\; \mathrm{tr} (F_{\mu\nu}\,^*F_{\mu\nu}) \right\} \nonumber\\
&=& \frac{1}{2} \frac{1}{(4\pi)^2} \ln \frac{M^2}{\mu^2}\int \mathrm{d}^4x\; \mathrm{tr} (F_{\mu\nu}\,^*F_{\mu\nu}). \label{DeltaGammaM}
\end{eqnarray}
Therefore, from (\ref{DeltaGammamM}) and (\ref{DeltaGammaM}), we can secure the general result
\begin{eqnarray}
\Delta\Gamma(A;m) = \frac{1}{2} \frac{1}{(4\pi)^2} \ln \frac{m^2}{\mu^2}\int \mathrm{d}^4x\; \mathrm{tr} (F_{\mu\nu}\,^*F_{\mu\nu}). \label{DeltaGammaresult}
\end{eqnarray}
Gauge invariance is manifest in this formula. Also notice the way how mass $m$ and the renormalization scale $\mu$ enters this formula.

Using the above finding with (\ref{GammaDeltaGamma1}) or (\ref{GammaDeltaGamma2}) yields the desired formula for the spinor effective action. That is, we can express our (minimally renormalized) spinor effective action by either
\begin{eqnarray}
\Gamma_{\text{ren}}^F(A;m) = 2 \Gamma_{\text{ren}}^{(+)}(A;m) - \frac{1}{2} \frac{1}{(4\pi)^2} \ln \frac{m^2}{\mu^2} \int \mathrm{d}^4x\; \mathrm{tr} (F_{\mu\nu}\,^*F_{\mu\nu}) \label{Gammaren1}
\end{eqnarray}
or
\begin{eqnarray}
\Gamma_{\text{ren}}^F(A;m) = 2 \Gamma_{\text{ren}}^{(-)}(A;m) + \frac{1}{2} \frac{1}{(4\pi)^2} \ln \frac{m^2}{\mu^2} \int \mathrm{d}^4x\; \mathrm{tr} (F_{\mu\nu}\,^*F_{\mu\nu}). \label{Gammaren2}
\end{eqnarray}
Thus, to determine $\Gamma_{\text{ren}}^F(A;m)$, it suffices to evaluate $\Gamma_{\text{ren}}^{(+)}(A;m)$ or $\Gamma_{\text{ren}}^{(-)}(A;m)$ (i.e., the effective action associated with a particular chiral projection of the second order Dirac operator (\ref{chiraldecomposition})), and not both. The calculational labor might be greatly reduced by choosing a particular chiral projection. In a self-dual gauge field background for instance, it is the amplitude $\Gamma_{\text{ren}}^{(+)}(A;m)$ that is simpler. This is because, with $F_{\mu\nu}=\,^*F_{\mu\nu}$, the quantity $F^{(+)}(s)$ defined in (\ref{Fdef}) becomes
\begin{eqnarray}
F^{(+)}(s) = 2 \int \mathrm{d}^4x\; \mathrm{tr} \left\langle x \left| \left[ e^{-s(-D^2)}-e^{-s(-\partial^2)} \right] \right| x \right\rangle \equiv 2F_{\text{scalar}},
\end{eqnarray}
and accordingly, for the amplitude $\Gamma_{\text{ren}}^{(+)}(A;m)$ (see (\ref{Gammarenpm})), we find
\begin{eqnarray}
\Gamma_{\text{ren}}^{(+)}(A;m) &=& \lim_{\Lambda\to\infty} \left\{ \int_0^\infty \frac{\mathrm{d}s}{s} \left( e^{-m^2s}-e^{-\Lambda^2s} \right) F_{\text{scalar}} + \frac{1}{12} \frac{1}{(4\pi)^2} \ln \frac{\Lambda^2}{\mu^2} \int \mathrm{d}^4x\; \mathrm{tr} \left( F_{\mu\nu}F_{\mu\nu} \right) \right\} \nonumber\\
&=& -\Gamma_{\text{ren}}^S(A;m), \label{sGamma}
\end{eqnarray}
viz., up to sign, $\Gamma_{\text{ren}}^{(+)}(A;m)$ coincides with the (minimally renormalized) scalar effective action for which various calculational methods have been developed. [On the other hand, the \emph{direct} evaluation of $\Gamma_{\text{ren}}^{(-)}(A;m)$ will be very nontrivial]. In the self-dual case, we also recover (\ref{fermionscalarrelation}) from our formula (\ref{Gammaren1}) since $n_F$, the number of normalizable zero modes of the operator $\slashed{D}$ in a self-dual background is dictated by the value $\frac{1}{(4\pi)^2} \int \mathrm{d}^4x\; \mathrm{tr} (F_{\mu\nu}\,^*F_{\mu\nu})$ \cite{bcl}.

We remark that, although Euclidean spacetime has been assumed for our development, there should exist formulas similar to (\ref{S1}) or (\ref{S2}) and (\ref{Gammaren1}) or (\ref{Gammaren2}) for the Minkowski-space (Feynman) propagator and effective action. This is because, in our derivations given above, the spacetime signature is in no way crucial. But, in the case of the formulas (\ref{Sform}) and (\ref{fermionscalarrelation}), we do not have the direct Minkowski analogues since the notion of the (anti-)self-duality for the field strengths is meaningful only in the Euclidean spacetime.

\section{The case of constant field strengths} \label{sec3}
As an important finding obtained in the previous section, we had the relations between $\Gamma^{F}_{\text{ren}}(A;m)$ and $\Gamma^{(\pm)}_{\text{ren}}(A;m)$, which are given in (\ref{Gammaren1}) or in (\ref{Gammaren2}). It was shown that one of these relations, (\ref{Gammaren1}), reduces to (\ref{fermionscalarrelation}) in the case of self-dual backgrounds. As an explicit check on the validity of those relations with general non-self-dual backgrounds, we here consider the case of constant field strengths for which both of the quantities $\Gamma^{(\pm)}_{\text{ren}}(A;m)$ can be evaluated in a closed form. The explicit forms of the effective action $\Gamma^{F}_{\text{ren}}(A;m)$ with constant $F_{\mu\nu}$ can be found for instance in Refs. \cite{schwinger, leutwyler}. However, each of  $\Gamma^{(\pm)}_{\text{ren}}(A;m)$ was not separately analyzed there, and so we will evaluate them explicitly in this section. In an Abelian gauge theory like QED, the constant field strength is defined by the conditions $\partial_\mu F_{\nu\lambda}=0$. On the other hand, in a non-Abelian theory, we have the covariant conditions $[D_\mu, F_{\nu \lambda}]=0$ instead. These conditions imply the commuting properties, $[F_{\mu \nu}, F_{\alpha \beta}]=0$. Hence we may diagonalize the whole system in the isospin space: then, the field strength $F_{\mu\nu}$ may be taken to be of a diagonal form $(F_{\mu\nu})_{ii} \delta_{ij}$ in the isospin space  ($i$, $j$ are isospin indices) for example. With this understanding, we will suppress these isospin indices for the notational simplicity below.

Let us begin with the consideration of the functions $F^{(\pm)}(s)$ in (\ref{Fdef}). In our case the nontrivial operator in (\ref{Fdef}) can be factorized as
\begin{equation}
e^{-s(-D^2-\frac{1}{2}\eta^{(\mp)}_{\mu\nu a}\sigma_a F_{\mu\nu})} = e^{-s(-D^2)} \cdot e^{\frac{s}{2}\eta^{(\mp)}_{\mu\nu a}\sigma_a F_{\mu\nu}}. \label{operatorfactorize}
\end{equation}
Note that the second factor carries spinor indices while the first does not. Since the total trace for the first factor is available in the literature \cite{schwinger,leutwyler}, we will concentrate on the second factor. Using the commuting properties of $F_{\mu\nu}$ and the well-known identity involving Pauli matrices
\begin{equation}
e^{\sigma_a v_a}= \cosh \sqrt{v_a v_a}
+\sigma_a v_a \frac{\sinh \sqrt{ v_a v_a}}{ \sqrt{v_a v_a}}
\end{equation}
(here, $v_a$ can be an arbitrary vector), one may perform the 2-component spinor trace for the second factor in (\ref{operatorfactorize}) as
\begin{eqnarray}
\mathrm{tr}_\mathrm{sp}  e^{\frac{s}{2}\eta^{(\mp)}_{\mu\nu a}\sigma_a F_{\mu\nu}}
&=&2 \cosh  \frac{s}{2}\sqrt{\eta^{(\mp)}_{\mu\nu a}\eta^{(\mp)}_{\alpha\beta a}
 F_{\mu\nu}F_{\alpha\beta}} \nonumber\\
&=& 2\cosh  \sqrt{2 s^2 (\mathcal{F}_1 \mp \mathcal{F}_2)}, \label{trsp}
\end{eqnarray}
where
\begin{equation}
\mathcal{F}_1 = \frac{1}{4}F_{\mu\nu}F_{\mu\nu}, \quad
\mathcal{F}_2 = \frac{1}{4}F_{\mu\nu} \,^{*}F_{\mu\nu}.  \label{FFdual}
\end{equation}
To derive the second form in (\ref{trsp}), we have made use of the identity
\begin{equation}
\eta^{(\mp)}_{\mu\nu a}\eta^{(\mp)}_{\alpha\beta a}
=\delta_{\mu\alpha}\delta_{\nu\beta}-\delta_{\nu\alpha}\delta_{\mu\beta}
\mp \epsilon_{\mu\nu\alpha\beta}. \label{etaid}
\end{equation}

Based on (\ref{trsp}) and the known result \cite{schwinger,leutwyler} on the trace for the first factor in (\ref{operatorfactorize}), we now have
\begin{equation}
\overline{\mathrm{Tr}} \left[e^{-s(-D^2-\frac{1}{2}\eta^{(\mp)}_{\mu\nu a}\sigma_a F_{\mu\nu})}\right]=
 \frac{2 V}{(4\pi s)^2}\; \mathrm{tr} \left(
 e^{-L}  \cosh\sqrt{2 s^2 (\mathcal{F}_1 \mp \mathcal{F}_2)}\right), \label{evolution}
\end{equation}
where $V$ is the volume of the spacetime, and
\begin{equation}
L= \frac{1}{2}\; \mathrm{tr}_\mathrm{L} \ln \left[( s\mathbf{F})^{-1} \sin (s\mathbf{F})\right].   \label{constL}
\end{equation}
In (\ref{constL}), `$\mathrm{tr}_\mathrm{L}$' denotes the trace over Lorentz indices and $\mathbf{F}$ is the matrix defined by the relation $[\mathbf{F}]_{\mu\nu}=F_{\mu\nu}$. This $4\times4$ matrix $\mathbf{F}$ has four eigenvalues $\pm i f_{+}$ and $\pm i f_{-}$, where
\begin{eqnarray}
f_\pm&=&\frac{1}{\sqrt{2}} \left( \sqrt{\mathcal{F}_1 +\mathcal{F}_2} \pm \sqrt{\mathcal{F}_1 -\mathcal{F}_2}\right)
\end{eqnarray}
with $\mathcal{F}_1$ and $\mathcal{F}_2$ defined in (\ref{FFdual}). Using these eigenvalues, we may express the factor $e^{-L}$ by the form
\begin{equation}
e^{-L}= \frac{s^2 f_+f_-}{\sinh(sf_+)\sinh(sf_-)}. \label{eL}
\end{equation}
We also simplify the factor $\sqrt{2s^2 (\mathcal{F}_1 \mp \mathcal{F}_2)}$ as
\begin{eqnarray}
\cosh s\sqrt{2 (\mathcal{F}_1 \mp \mathcal{F}_2)}&=& \cosh \left( sf_+\mp sf_-\right) \nonumber\\
&=&\cosh( sf_+) \cosh( sf_-) \mp \sinh( sf_+ ) \sinh( sf_-). \label{coshs}
\end{eqnarray}
Plugging (\ref{eL}) and (\ref{coshs}) into (\ref{evolution}), we find the following forms for the functions $F^{(\pm)}(s)$:
\begin{equation}
F^{(\pm)}(s)=\frac{2V}{(4\pi s)^2}\; \mathrm{tr}\left[s^2 f_{+}f_{-}
\left(\coth (sf_+) \coth (sf_-) \mp 1 \right)  -1\right]. \label{Fpm}
\end{equation}

From (\ref{Fpm}) we find $F(s)$, the sum of $F^{(+)}(s)$ and $F^{(-)}(s)$, to be
\begin{equation}
F(s)=\frac{4V}{(4\pi )^2}\; \mathrm{tr}\left[ f_{+} f_{-}
\coth (sf_+) \coth (sf_-)   -\frac{1}{s^2}\right].
\end{equation}
This is  precisely the expression appropriate for a Dirac fermion in \cite{schwinger,leutwyler}, and using this in (\ref{Gammaren}) yields the known result for the renormalized effective action $\Gamma_{\text{ren}}(A;m)$. On the other hand, according to (\ref{Fpm}), the difference between $F^{(+)}(s)$ and $F^{(-)}(s)$ is given by
\begin{equation}
F^{(+)}(s)-F^{(-)}(s)=-\frac{4V}{(4\pi)^2}\; \mathrm{tr}\mathcal{F}_2.
\end{equation}
Using this result in (\ref{DeltaGammadef}), we can determine the quantity $\Delta\Gamma(A;m)$ (with constant field strengths) as
\begin{eqnarray}
\Delta\Gamma(A;m)&=&\frac{1}{2}
\frac{V}{(4\pi)^2}  \lim_{\Lambda\to\infty}  \left\{
\int_0^\infty \frac{\mathrm{d}s}{s} \left( e^{-m^2 s} -e^{-\Lambda^2 s} \right) (-4\;\mathrm{tr}\mathcal{F}_2)
 +  \ln\frac{\Lambda^2}{\mu^2}\;\mathrm{tr}(F_{\mu\nu}\,^*F_{\mu\nu}) \right\} \nonumber \\
&=&\frac{1}{2}
\frac{V}{(4\pi)^2}
\ln\frac{m^2}{\mu^2}\;\mathrm{tr}(F_{\mu\nu}\,^*F_{\mu\nu}).
\end{eqnarray}
This confirms our general formula (\ref{DeltaGammaresult}), and then from (\ref{GammaDeltaGamma1}) and (\ref{GammaDeltaGamma2}) the relations (\ref{Gammaren1}) and (\ref{Gammaren2}) are immediate.

\section{Concluding remarks} \label{sec4}
For a general massive Dirac field in arbitrary gauge field background, we have derived appropriate formulas which make possible the construction of the full Dirac propagator and spinor effective action from the information on those pertaining to a particular chirality sector of the second-order Dirac operator $\slashed{D}\slashed{D}+m^2$. This will serve useful purpose when one wishes to calculate the spinor effective action in some given background fields. Assuming an SU(2) gauge theory, take for instance gauge fields of the form $A_\mu (x) = \eta_{\mu\nu a}^{(+)} x_\nu \phi_a(r)$, where $r \equiv \sqrt{x_\mu x_\mu}$ and $\phi_a(r)$ is a certain $2 \times 2$ matrix function of $r$ (e.g., $\phi_a(r) = f(r) \tau_a$ or $\phi_a(r) = g(r) \tau_3 \delta_{a3}$, $\tau_a$ being $2 \times 2$ Pauli matrices). Then we find a relatively simple expression for $\eta_{\mu\nu a}^{(+)} F_{\mu\nu}$
\begin{eqnarray}
\eta_{\mu\nu a}^{(+)} F_{\mu\nu} = -8\phi_a(r) - 2r\partial_r \phi_a(r) - ir^2\epsilon_{abc} \left[ \phi_b(r), \phi_c(r) \right],
\end{eqnarray}
and so it will be possible to study $\Gamma_{\text{ren}}^{(-)}(A;m)$ by using, say, the method developed in Refs. \cite{insdet,radial}; in contrast, due to the complicated (non-radial) expression of $\eta_{\mu\nu a}^{(-)} F_{\mu\nu}$, no such analysis is readily available for $\Gamma_{\text{ren}}^{(+)}(A;m)$. In this case, we can use our formula (\ref{GammaDeltaGamma2}) to determine $\Gamma_{\text{ren}}(A;m)$, i.e., the analysis for $\Gamma_{\text{ren}}^{(+)}(A;m)$ can be avoided completely.
The spinor effective action in various form of Abelian gauge fields with radial symmetry has been calculated following this strategy and it will be reported separately \cite{abelianspinor}.

Also recall that, in an (anti-)self-dual gauge field background, we had formulas relating the vector (i.e., spin-1) propagator and functional determinant to those of a scalar field in the same background \cite{bccl,brownlee,thooft,browncreamer,pac}. Then, as in the case of a spin-$\frac{1}{2}$ field, can one have a suitable generalization of these results so that they become valid also with a general background? It appears not to be the case, as the (massless) vector fluctuation operator in a general background does not possess a simple factorizable structure of the (massless) Dirac operator.

\section*{Acknowledgments}
We are grateful to G. Dunne for helpful discussions. This work was supported by the Basic Science Research Program through the National Research Foundation of Korea (NRF) funded by the Ministry of Education, Science and Technology( No. 2009-0076297(C.L.) and No. 2010-0011223  (H.M.)).

\newpage
\appendix
\section{Renormalization counterterms and the heat kernel expansion} \label{appendixA}
The Schwinger proper-time method is particularly expedient in finding renormlaization counterterms, which are related to regulator-dependent contributions arising from the small-$s$ end of the proper-time integral for the effective action. For such study it is useful to consider the proper-time Green function (or the heat kernel)
\begin{eqnarray}
&& \langle xs | y \rangle \equiv \left\langle x \left| e^{-s\hat{M}} \right| y \right\rangle, \nonumber\\
&& \hat{M} = -D_\mu D_\mu + Q,
\end{eqnarray}
($Q$ refers to a generic non-derivative operator), which satisfies the heat equation
\begin{eqnarray}
-\frac{\partial}{\partial s} \langle xs | y \rangle = \left[ -D_\mu D_\mu + Q \right] \langle xs | y \rangle, \qquad (s>0). \label{heateq}
\end{eqnarray}
The function $\langle xs | y \rangle$ admits a small-$s$ expansion
\begin{eqnarray}
s\to0+ \;:\; \langle xs | y \rangle = \frac{1}{(4\pi s)^2} e^{-\frac{(x-y)^2}{4s}} \left\{ a_0(x,y) + a_1(x,y) s + a_2(x,y) s^2 + \cdots \right\} \label{heatkernelexpansion}
\end{eqnarray}
with $a_0(x,x)=1$. Then looking at the proper-time integral such as the one in (\ref{properGamma}), one will recognize that the coincidence limits $a_1(x,x)$ and $a_2(x,x)$ are entirely responsible for the renormalization counterterms. The heat kernel expansion (\ref{heatkernelexpansion}) will also be useful in deriving the large mass expansion of the effective action.

To determine the above coincidence limits, one may insert the series (\ref{heatkernelexpansion}) into (\ref{heateq}), to obtain
\begin{eqnarray}
O(s^{-1}) &:& (x-y)_\mu D_\mu a_0 = 0, \\
O(s^0) &:& -a_1 = (x-y)_\mu D_\mu a_1 + \left[ -D^2+Q \right] a_0, \\
O(s^1) &:& -2a_2 = (x-y)_\mu D_\mu a_2 + \left[ -D^2+Q \right] a_1,
\end{eqnarray}
etc. Then, through considering successive differentiations of these relations (using $D$'s!) before taking the coincidence limits $x=y$, one can easily deduce that
\begin{eqnarray}
a_1(x,x) &=& -Q(x), \label{a1}\\
a_2(x,x) &=& \frac{1}{12} [D_\mu,D_\nu] [D_\mu,D_\nu] - \frac{1}{6} [D_\mu,[D_\mu,Q]] + \frac{1}{2} Q^2(x). \label{a2}
\end{eqnarray}

The result of applying the above findings to some operators $\hat{M}$ relevant in this work is as follows.
\begin{enumerate}[(i)]
  \item $\hat{M}=-D^2$ (scalar): here, we have $Q=0$ and so
  \begin{eqnarray}
  a_1(x,x)=0, \qquad a_2(x,x) = \frac{1}{12} [D_\mu,D_\nu] [D_\mu,D_\nu] = -\frac{1}{12} F_{\mu\nu}F_{\mu\nu}.
  \end{eqnarray}
  These results explain the renormalization counter term appearing in (\ref{sGamma}).
  \item $\hat{M} = \slashed{D}\slashed{D} = -D^2+\frac{1}{2}\sigma_{\mu\nu}F_{\mu\nu}$ (Dirac spinor): here, we have $Q=\frac{1}{2}\sigma_{\mu\nu}F_{\mu\nu}$ and then, using our formulas (\ref{a1}) and (\ref{a2}) and taking the trace,
  \begin{eqnarray}
  \mathcal{T}\!\mathit{r}\; a_1(x,x) = 0, \qquad \mathcal{T}\!\mathit{r}\; a_2(x,x) = \frac{2}{3} \mathrm{tr}\left( F_{\mu\nu}F_{\mu\nu} \right).
  \end{eqnarray}
  These explain the renormalization counterterm appearing in (\ref{Gammaren}).
  \item $\hat{M} = -D^2-\frac{1}{2}\eta_{\mu\nu a}^{(\mp)}\sigma_a F_{\mu\nu}$ (chirally projected spinor): here, we have $Q=-\frac{1}{2}\eta_{\mu\nu a}^{(\mp)}\sigma_a F_{\mu\nu}$ and then, using our formulas (\ref{a1}) and  (\ref{a2}) with the identity (\ref{etaid}),
  we find the traces
  \begin{eqnarray}
  \overline{\mathcal{T}\!\mathit{r}}\; a_1(x,x) = 0, \qquad \overline{\mathcal{T}\!\mathit{r}}\; a_2(x,x) = \frac{1}{3} \mathrm{tr}(F_{\mu\nu}F_{\mu\nu}) \mp \frac{1}{2} \mathrm{tr}(F_{\mu\nu}\,^*F_{\mu\nu}). \label{heatkernelresult3}
  \end{eqnarray}
  These results explain the renormalization counterterms appearing in (\ref{Gammarenpm}) and were also used to derive our large mass expansion result given in (\ref{DeltaGammaM}).
\end{enumerate}

\section{Derivation of the formula (\ref{chiralanomaly}) by the proper time method} \label{appendixB}
As for the quantity given by
\begin{eqnarray}
I = \bar{m}^2 \frac{\mathrm{d}(\Delta\Gamma(A;\bar{m}))}{\mathrm{d}\bar{m}^2},
\end{eqnarray}
we may utilize our form (\ref{DeltaGammap}) to write this as
\begin{eqnarray}
I &=& -\frac{1}{2} \int_0^\infty \mathrm{d}s\; \bar{m}^2 e^{-\bar{m}^2s}\; \overline{\mathrm{Tr}} \left[ e^{-s(-\sigma \cdot D \; \bar{\sigma } \cdot D)} - e^{-s(-\bar{\sigma} \cdot D \; \sigma \cdot D)} \right] \nonumber\\
&=& -\frac{1}{2} \int_0^\infty \mathrm{d}s\; \bar{m}^2 e^{-\bar{m}^2s}\; \mathrm{Tr} \left[ \gamma_5 e^{-s\slashed{D}\slashed{D}} \right]. \label{I2}
\end{eqnarray}
We shall here show via manipulating this expression that this quantity in in fact equal to $\frac{1}{32\pi^2} \int \mathrm{d}^4x\; \mathrm{tr}(F_{\mu\nu}\,^*F_{\mu\nu})$, for any value of $\bar{m}^2$. Using $\bar{m}^2e^{-\bar{m}^2s} = -\frac{\partial}{\partial s} e^{-\bar{m}^2s}$ and integrating by parts, (\ref{I2}) can be recast as
\begin{eqnarray}
I = \frac{1}{2} e^{-\bar{m}^2s}\; \mathrm{Tr} \left. \left[ \gamma_5 e^{-s\slashed{D}\slashed{D}} \right] \right|_{s=0}^{s=\infty} - \frac{1}{2} \int_0^\infty \mathrm{d}s\; e^{-\bar{m}^2s}\; \mathrm{Tr} \left[ \gamma_5 \frac{\partial}{\partial s} e^{-s\slashed{D}\slashed{D}} \right]. \label{I3}
\end{eqnarray}
The second term in the right hand side of (\ref{I3}) actually vanishes. To show this, notice that
\begin{eqnarray}
&& \mathrm{Tr} \left[ \gamma_5 \frac{\partial}{\partial s} e^{-s\slashed{D}\slashed{D}} \right] = \int \mathrm{d}^4x\; \mathcal{T}\!\mathit{r} \left[ \gamma_5 \frac{\partial}{\partial s} \left\langle x \left| e^{-s\slashed{D}\slashed{D}} \right| x \right\rangle \right] \nonumber\\
&&\qquad = -\frac{1}{2} \int \mathrm{d}^4x\; \mathcal{T}\!\mathit{r} \left. \left[ \gamma_5 \slashed{D}^{(x)} \slashed{D}^{(x)} \left\langle x \left| e^{-s\slashed{D}\slashed{D}} \right| y \right\rangle + \gamma_5 \slashed{D}^{(x)} \left\langle x \left| e^{-s\slashed{D}\slashed{D}} \right| y \right\rangle \overleftarrow{\slashed{D}}^{(y)} \right] \right|_{y=x},
\end{eqnarray}
since $\langle x | e^{-s\slashed{D}\slashed{D}} | y \rangle$ satisfies the heat equation
\begin{eqnarray}
-\frac{\partial}{\partial s} \left\langle x \left| e^{-s\slashed{D}\slashed{D}} \right| y \right\rangle &=& \slashed{D}^{(x)} \slashed{D}^{(x)} \left\langle x \left| e^{-s\slashed{D}\slashed{D}} \right| y \right\rangle \nonumber\\
 &=& \slashed{D}^{(x)} \left\langle x \left| e^{-s\slashed{D}\slashed{D}} \right| y \right\rangle \overleftarrow{\slashed{D}}^{(y)}.
\end{eqnarray}
Then, because of the relation
\begin{eqnarray}
&& \partial_\mu^{(x)} \mathcal{T}\!\mathit{r} \left[ \gamma_5 \gamma_\mu \left. \left\{ \slashed{D}^{(x)} \left\langle x \left| e^{-s\slashed{D}\slashed{D}} \right| y \right\rangle \right\} \right|_{y=x} \right] \nonumber\\
&&\qquad =  \lim_{y\to x} \mathrm{tr} \left\{ \gamma_5 \slashed{D}^{(x)} \slashed{D}^{(x)} \left\langle x \left| e^{-s\slashed{D}\slashed{D}} \right| y \right\rangle + \gamma_5 \slashed{D}^{(x)} \left\langle x \left| e^{-s\slashed{D}\slashed{D}} \right| y \right\rangle \overleftarrow{\slashed{D}}^{(y)} \right\},
\end{eqnarray}
we can further write
\begin{eqnarray}
\mathrm{Tr} \left[ \gamma_5 \frac{\partial}{\partial s} e^{-s\slashed{D}\slashed{D}} \right] = -\frac{1}{2} \int \mathrm{d}^4x\; \partial_\mu^{(x)} \mathcal{T}\!\mathit{r} \left[ \gamma_5 \gamma_\mu \left. \left\{ \slashed{D}^{(x)} \left\langle x \left| e^{-s\slashed{D}\slashed{D}} \right| y \right\rangle \right\} \right|_{y=x} \right]
\end{eqnarray}
and hence the very second term in (\ref{I3}) becomes
\begin{eqnarray}
&& -\frac{1}{2} \int_0^\infty \mathrm{d}s\; e^{-\bar{m}^2s}\; \mathrm{Tr} \left[ \gamma_5 \frac{\partial}{\partial s} e^{-s\slashed{D}\slashed{D}} \right] \nonumber\\
&&\qquad = \frac{1}{4} \oint_{|x|\to \infty } \mathrm{d}S^\mu \int_0^\infty \mathrm{d}s\; \mathcal{T}\!\mathit{r} \left[ \gamma_5 \gamma_\mu \left. \left\{ \slashed{D}^{(x)} \left\langle x \left| e^{-s(\slashed{D}\slashed{D}+\bar{m}^2)} \right| y \right\rangle \right\} \right|_{y=x} \right] \nonumber\\
&&\qquad = 0. \label{secondtermvanishes}
\end{eqnarray}
[Notice that what appears in the integrand of the surface integral here is essentially the vacuum expectation value of the axial vector current $\psi^\dagger \gamma_5 \gamma_\mu \psi$].

Thanks to (\ref{secondtermvanishes}), we now have
\begin{eqnarray}
I &=& -\frac{1}{2} \lim_{s\to 0+} \mathrm{Tr} \left[ \gamma_5 e^{-s\slashed{D}\slashed{D}} \right] \nonumber\\
&=& -\frac{1}{2} \lim_{s\to 0+} \overline{\mathrm{Tr}} \left[ e^{-s(-\sigma \cdot D \; \bar{\sigma } \cdot D)} - e^{-s(-\bar{\sigma} \cdot D \; \sigma \cdot D)} \right]. \label{Ilimit}
\end{eqnarray}
The limit (\ref{Ilimit}) can readily be evaluated with the help of the heat kernel expansions given in Appendix \ref{appendixA} (see (\ref{heatkernelexpansion}) and (\ref{heatkernelresult3})) and the result is
\begin{eqnarray}
I &=& -\frac{1}{2} \lim_{s\to 0+} \frac{1}{(4\pi s)^2} \left\{ s^2 \cdot \left(-\frac{1}{2}\right) \cdot 2 \int \mathrm{d}^4x\; \mathrm{tr} (F_{\mu\nu}\,^*F_{\mu\nu}) + O(s^3) \right\} \nonumber\\
&=& \frac{1}{32\pi^2} \int \mathrm{d}^4x\; \mathrm{tr} (F_{\mu\nu}\,^*F_{\mu\nu}).
\end{eqnarray}


\begin{thebibliography}{12345}
\bibitem{bccl} L. S. Brown, R. D. Carlitz, D. B. Creamer, and C. Lee, Phys. Rev. D \textbf{17}, 1583 (1978); Phys. Lett. \textbf{B70}, 180 (1977).
\bibitem{brownlee} L. S. Brown and C. Lee, Phys. Rev. D \textbf{18}, 2180 (1978).
\bibitem{thooft} G. 't Hooft, Phys. Rev. D \textbf{14}, 3432 (1976); \textbf{18}, 2199(E) (1978).
\bibitem{browncreamer} L. S. Brown and D. B. Creamer, Phys. Rev. D \textbf{18}, 3695 (1978).
\bibitem{pac} C. Lee, H. W. Lee, and P. Y. Pac, Nucl. Phys. \textbf{B201}, 429 (1982).
\bibitem{insdet} G. V. Dunne, J. Hur, C. Lee and H. Min, Phys. Rev. Lett. \textbf{94}, 072001 (2005); Phys. Rev. D \textbf{71}, 085019 (2005).
\bibitem{zumino} B. Zumino, Phys. Lett. \textbf{B69}, 369 (1977); A. D'Adda and P. Di Vecchia, Phys. Lett. \textbf{B73}, 162 (1978).
\bibitem{adler} S. Adler, Phys. Rev. \textbf{177}, 2426 (1969); J. S. Bell and R. Jackiw, Nuovo Cimento \textbf{60A}, 49 (1969).
\bibitem{schwinger} J. Schwinger, Phys. Rev. \textbf{82}, 664 (1951).
\bibitem{kwon} O.K. Kwon, C. Lee, and H. Min, Phys. Rev. D \textbf{62}, 114022 (2000).
\bibitem{Fry} M.P. Fry, Phys. Rev. D \textbf{75}, 065002 (2007), arXiv:hep-th/0612218; Phys. Rev. D \textbf{81}, 107701 (2010), arXiv:hep-th/1005.4849. 
\bibitem{bcl} L. S. Brown, R. D. Carlitz, and C. Lee, Phys. Rev. D \textbf{16}, 417 (1977); See also J. Kiskis, Phys. Rev. D \textbf{15}, 2329 (1977).
\bibitem{weinberg} E.J. Weinberg, Phys. Rev. D \textbf{24}, 2669 (1981).
\bibitem{leutwyler} H. Leutwyler, Nucl. Phys. \textbf{B179}, 129 (1981).
For a review, see: G.V. Dunne, arXiv:hep-th/0406216, in ``Ian Kogan Memorial Collection; From
Fields to Strings: Circumnavigating Theoretical Physics'', M. Shifman (World Scientific, Singapore, 2004), Vol. 1, p. 445.
\bibitem{radial} G.V. Dunne, J. Hur, C. Lee, and H. Min, Phys. Rev. \textbf{77}, 045004 (2008).
\bibitem{abelianspinor} G.V. Dunne, A. Huet, J. Hur, and H. Min, ``Renormalized Spinor Effective Action in Radially Symmetric Backgrounds'' , in preparation. 
\end{thebibliography}
\end{document}